\documentclass[9pt,conference]{IEEEtran}

\usepackage{cite}
\usepackage{amsmath,amssymb,amsfonts}
\usepackage{algorithmic}
\usepackage{graphicx}
\usepackage{textcomp}
\usepackage{booktabs}
\usepackage{localmath}
\usepackage{matvecsymb}
\usepackage{siunitx}
\usepackage{enumitem}

\usepackage[preprint]{waspaa25}

\usepackage{bm} %

\title{Source Separation by Flow Matching}

\name{Robin Scheibler, John R. Hershey, Arnaud Doucet, and Henry Li}
\address{Google DeepMind \\
robinsch@google.com
}

\begin{document}

\newcommand{\rmd}{\textup{d}}
\newcommand{\lossdb}{\calL_{\operatorname{dB}}}

\maketitle

\begin{abstract}
We consider the problem of single-channel audio source separation with the goal of reconstructing $K$ sources from their mixture.
We address this ill-posed problem with FLOSS (FLOw matching for Source Separation), a constrained generation method based on flow matching, ensuring strict mixture consistency.
Flow matching is a general methodology that, when given samples from two probability distributions defined on the same space, learns an ordinary differential equation to output a sample from one of the distributions when provided with a sample from the other.
In our context, we have access to samples from the joint distribution of $K$ sources and so the corresponding samples from the lower-dimensional distribution of their mixture.
To apply flow matching, we augment these mixture samples with artificial noise components to 
match the dimensionality of the $K$ source distribution.
Additionally, as any permutation of the sources yields the same mixture, we adopt an equivariant formulation of flow matching which relies on a 
neural network architecture that is equivariant by design.
We demonstrate the performance of the method for the separation of overlapping speech.
\end{abstract}


\section{Introduction and Contributions}

The goal of \emph{source separation} techniques is to recover individual signals from recordings of mixtures of them. Given its critical applications in domains such as audio, this has been an active research area since the mid-1990s~\cite{makinoAudioSourceSeparation2018}. 

If multiple mixtures of the same signals are available, the problem can be well-posed and is often amenable to multi-channel methods such as independent component analysis~\cite{makinoAudioSourceSeparation2018}.
In contrast, the extremely under-determined case of a single microphone is much more challenging; early attempts based on non-negative matrix factorization had limited success \cite{smaragdis_static_2014}.
Deep learning approaches represented a breakthrough. Deep clustering introduced a learned embedding for time-frequency bins of the short-time Fourier transform (STFT), allowing them to be clustered by source and enabling separation by masking ~\cite{hersheyDeepClusteringDiscriminative2016}.
More recently, end-to-end permutation invariant training (PIT)~\cite{kolbaekMultitalkerSpeechSeparation2017} has been used to train supervised separation networks such as Conv-TasNet~\cite{luo_conv-tasnet_2019}, Band-Split RNN~\cite{yu_high_2023}, Mel-band-split RoFormer~\cite{Wang2023-mc}, and TF-Locoformer~\cite{Saijo2024-rq}.

While these regression-based techniques have been wildly successful, they also suffer from typical regression artifacts~\cite{larsen2016autoencoding}, and cannot recover signals buried in strong interference.  Generative models are well positioned to avoid such artifacts by in-painting over the missing parts \cite{subakan_generative_2018,jayaram2020source}.  Here we focus on diffusion and flow techniques.

Diffusion models, introduced by \cite{sohl2015deep} and further developed by \cite{ho2020denoising,song2020score}, exhibit state-of-the-art performance in many domains including speech \cite{Chen2021-sh,lee_priorgrad_2022,koizumi_specgrad_2022,Lemercier2024-yv}.
In their original form, a data distribution is transformed into a noise distribution in order to learn a reversal of this transformation, mapping noise samples to data samples.
Such methodology can also be extended to conditional simulation by introducing a suitable guidance term \cite{song2020score,ho2022classifier}.
Diffusion-based sound separation has been approached in various ways, such as using a mixture conditional diffusion in the mel-spectral domain~\cite{chen_sepdiff_2023}, or using a multi-source time-domain diffusion model with a guidance term promoting mixture consistency~\cite{mariani2023multi}, as well as for post-processing of conventional regression-based separation models ~\cite{wang2024noise}.
Alternatively, \cite{scheibler_diffusion-based_2023}~relies on a non-standard noising dynamic such that the mean of the diffusion process goes from the $K$ clean sources to $K$ copies of the mixture; see \cite{dong_edsep_2025} for an improved version. 

In this work, we build on recent extensions of diffusion models that learn maps between any two distributions defined on the same space \cite{peluchettinon2021,De-Bortoli2021-zy,lipman2022flow,albergo2023stochastic,liu2022let,Wang2022-ig}.
In particular, we focus on the popular \emph{flow matching}~\cite{lipman2022flow} methods, which define such maps through an ordinary differential equation (ODE) whose drift is learned by minimizing a simple regression objective.
We propose FLOSS, a generative source separation method using a non-standard modification of flow matching to accommodate the unique characteristics of the problem.
First, we introduce an interpolating path between the sources and mixture distributions that preserves their common subspace, restricting noise to an orthogonal subspace.
Second, to handle the permutation symmetry of source separation, we adapt the equivariant flow matching procedure of \cite{klein2024equivariant}. For this purpose we introduce a new permutation equivariant network architecture specialized for audio diffusion.
Separated speech samples are available at \url{https://google.github.io/df-conformer/floss/}.

\section{Background}

\subsection{Problem Formulation and Notation}

\begin{figure}
    \centering
    \includegraphics[width=0.8\linewidth]{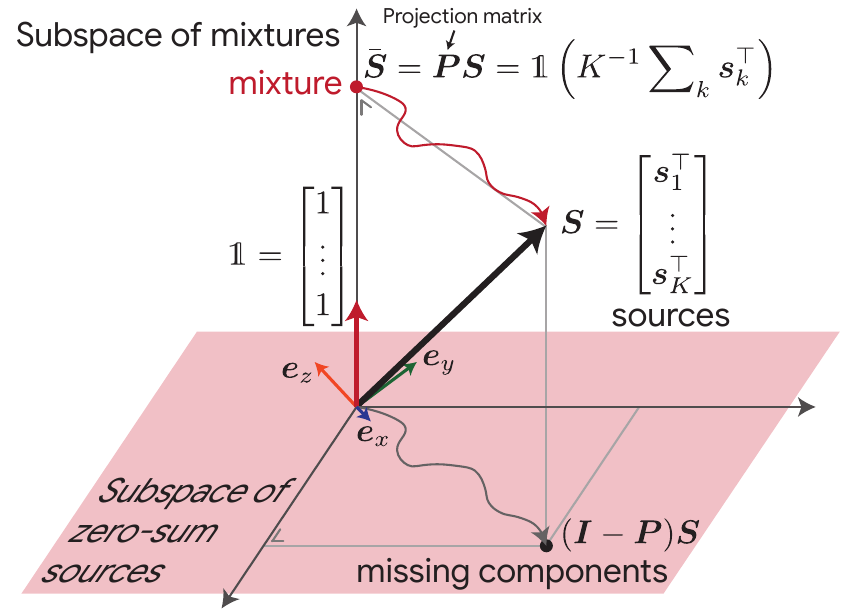}
    \caption{Illustration of the geometry of sources. The mixture lives in the uni-dimensional subspace spanned by $\vone = [1,\ldots,1]^\top$, while the missing components are in the sub-space of zero-sum signals.
    A possible path between the mixture and the separated sources is illustrated in red.
    The space has been rotated so that $\vone$ points up.}
    \label{fig:geometry}
\end{figure}

Let $\vs_{1:K}=(\vs_1,...,\vs_K)$ represent $K$ audio sources of length $L$.
We do not have access to these sources but only to the mixture defined by the forward model, $\vy = \sum_{k=1}^K \vs_k$.
Inferring the sources is thus a linear inverse problem.
To match the source dimensions and scale, we use the average of sources $\bar{\vs}=\vy^\top / K$,  defining the $ K\times L$ matrices of separate sources, and their average stacked $K$ times, as
\begin{equation}
    \mS=[\vs_1,...,\vs_K]^{\top},\quad \bar{\mS}=[\bar{\vs},...,\bar{\vs}]^{\top},
\end{equation}
which sum to the mixture, $\vy = \mS^\top\vone = \bar{\mS}^\top\vone$ where $\vone = [1,...,1]^\top$ is the all-ones vector of size $K$. We define the
projection matrices,
\begin{equation}
    \mP=\vone \vone^\top/K,\quad \mP^{\perp}=\mI_K-\mP,
\end{equation}
where $\mI_K$ is the identity matrix of size $K\times K$.  
The \emph{centering matrix}, $\mP^{\perp}$, removes the mean from a vector.
Conversely, $\mP$ projects onto the subspace orthogonal to $\mP^{\perp}$ spanned by $\vone$.  The forward model then becomes $\bar{\mS}=\mP\mS$.  See Fig.~\ref{fig:geometry} for an illustration.

\subsection{Diffusion Models and Source Separation}
Given sample $\vx_0 \in \mathbb{R}^d$ from some data distribution $p_0$, diffusion models consider the stochastic differential equation
\begin{equation}\label{eq:SDEnoise}
    \rmd \vx_t=f_t \vx_t\, \rmd t + g_t\, \rmd \vw_t,
\end{equation}
where $f_{t}$ is the \textit{drift}, $g_t$ is the \textit{diffusion} coefficient, and $\vw_t$ is a Wiener process. For suitable $f_t,g_t$, the distribution $p_t$ of $\vx_t$ is such that $p_1$ follows (approximately) a standard normal distribution $\mathcal{N}(0,\mathbf{I}_d)$ \cite{song2020score}. Thus the time-reversal of this process initialized at $t=1$ using a normal sample outputs a sample at $t=0$ from the data distribution and can be shown to satisfy
\begin{equation}\label{eq:SDEgen}
    \rmd \vx_t=(f_{t} \vx_t-g^2_{t} \nabla \log p_{t}(\vx_t))\, \rmd t + g_t\, \rmd \bar{\vw}_t,
\end{equation}
for $\bar{\vw}_t$ a standard Wiener process when time flows backward and $\rmd t$ is an infinitesimal negative timestep. 
The score $\nabla \log p_t$ is intractable but can be approximated by neural networks with a regression objective.

Sepdiff ~\cite{chen_sepdiff_2023} introduced a straightforward application of conditional diffusion for separation, in which the $\vx_0$ represents sources (e.g., $\vx_0 = \mS$) and their distribution is conditioned on the mixture $\vy$, so that  $\nabla \log p_{t}(\vx_t | \vy)$ is  approximated with a neural network. In ~\cite{mariani2023multi}, this is computed using $p_{t}(\vx_t | \vy) \propto p_{t}(\vx_t ) p_{t}(\vy | \vx_t)$, where $p_{t}(\vy | \vx_t)$ implements a likelihood term ensuring that the sources sum to the observed mixture, as in \cite{jayaram2020source}.  Inference then proceeds in the manner of diffusion posterior sampling \cite{chung2023diffusion}.

Diffsep~\cite{scheibler_diffusion-based_2023} introduced the following modified noising diffusion 
\begin{align}
\rmd \vx_t = -\gamma \mP^\perp \vx_t\,\rmd t + g_t\, \rmd \vw_t,
\label{eq:diffsep_sde}
\end{align}
such that when initialized from the $K$ sources, i.e., $\vx_0 = \mS$, the process induces the conditional distribution
$\vx_t|\vx_0 \sim \mathcal{N}(\vmu_t,\mSigma_t)$ where $\vmu_t = \bar{\mS} + e^{-\gamma t} \mP^\perp \mS$ such that $\mathbf{\mu}_1 \approx \bar{\mS}$.
EDsep~\cite{dong_edsep_2025} improves this method with a $\vmu_t$ prediction network and a better SDE sampler.

\subsection{Flow Matching}\label{subsec:flowmatching}

We consider the following variant of flow matching \cite{lipman2022flow,tong2024}.
Given a conditioning variable $\vc$, let $\vx_0 \sim p_0(\cdot |\vc), \vx_1 \sim p_1(\cdot |\vc)$ and define
\begin{equation}\label{eq:linearinterpolx_t}
     \vx_t(\vx_0, \vx_1) = t \vx_1 + (1-t) \vx_0, \quad \text{for}\ t\in(0, 1).
\end{equation}
The distribution $p_t(\cdot|\vc)$ of $\vx_t$ smoothly interpolates between $p_0(\cdot|\vc)$ and $p_1(\cdot|\vc)$. Flow matching learns the drift $v^\theta(t,\vx,\vc)$ of an ODE
\begin{equation}
    \rmd \vx_t=v^\theta(t,\vx_t,\vc) \rmd t,\quad \vx_0 \sim p_0(\cdot|\vc),
\end{equation}
such that $\vx_t \sim p_t(\cdot|\vc)$ for $t\in (0,1]$, in particular $\vx_1 \sim p_1(\cdot|\vc)$. 
This drift is obtained by minimizing the loss 
\begin{equation}
    \mathcal{L}(\theta)=\int_0^1 \mathbb{E}[||v^\theta(t,\vx_t(\vx_0,\vx_1),\vc)-(\vx_1-\vx_0)||^2]\,\rmd t,
    \label{eq:condflowmatchingloss}
\end{equation}
where the expectation is with respect to the data distributions $\vc \sim p(\vc), \vx_0 \sim p_0(\vx_0|\vc), \vx_1 \sim p_1(\vx_1|\vc)$.  A weighted version of this loss can also be used with time-varying weights.

\section{Flow matching for Source Separation}
Given access to source samples $\mS^i\sim p(\mS)$, our objective is to sample from the conditional distribution $p(\mS\,|\,\bar{\vs})$ for a mixture $\bar{\vs}$.

\subsection{Setup}
In our context, the conditioning variable for flow matching is $\vc=\bar{\vs}$ and $p_1(\vx_1|\vc)=p(\mS\,|\,\bar{\vs})$ is the ``target" conditional distribution. We also define an initial conditional distribution $p_0(\vx_0|\vc)$ via 
\begin{align}
    \vx_0 = \bar{\mS} + \mP^{\perp}\mZ,
    \label{eq:p(x_0|c)}
\end{align}
where $\mZ \sim \calN(\vzero, \mSigma_{\mZ})$; i.e. we fill the subspace orthogonal to $\vone$ with noise.
The choice of $\mSigma_{\mZ}$ is described in Section~\ref{sec:noise_shaping}.
In this notation, the flow matching objective \eqref{eq:condflowmatchingloss} becomes
\begin{equation}
    \mathcal{L}(\theta)=\int_0^1 \mathbb{E}[||v^\theta(t,\vx_t,\bar{\vs})-(\vx_1-\vx_0)||^2]\,\rmd t,
    \label{eq:condflowmatchinglossSourceSep}
\end{equation}
where $\vx_t$ satisfies \eqref{eq:linearinterpolx_t} and the expectation is with respect to $\bar{\vs} \sim p(\bar{\vs}), \vx_0 \sim p_0(\vx_0|\bar{\vs}), \vx_1 \sim p_1(\vx_1|\bar{\vs})$.
Furthermore, the target, $\vx_1 - \vx_0$, and $\vx_t$ have the following explicit expressions in \eqref{eq:condflowmatchinglossSourceSep}, respectively,
\begin{align}
    \vx_1-\vx_0 & =\mP^\perp (\mS-\mZ), \\
    \vx_t & =\bar{\mS}+\mP^\perp (t\mS+(1-t)\mZ).
\end{align}
Thus, the neural network $v^\theta(\,.\,)$ is trained to approximate
\begin{align}
v^{\textup{opt}}(t,\vx_t,\bar{\vs}) = \mP^\perp \mathbb{E}[\mS-\mZ|\vx_t,\bar{\vs}],
\end{align}
which suggests the following parameterization,
\begin{equation}\label{eq:paramdrift}
    v^\theta(t,\vx_t,\bar{\vs})=\mP^\perp \tilde{v}^\theta(t,\mP^\perp\vx_t,\bar{\vs}).
\end{equation}
By construction, an ODE with a learned drift \eqref{eq:paramdrift} initialized at $\vx_0$ in \eqref{eq:p(x_0|c)} is such that $\mP\vx_t=\bar{\mS}$ at each time $t$, and in particular $\vx_1$ is indeed such that the average of its component equals $\bar{\vs}$.
However, this approach does not take into account the fact that $p(\mS|\bar{\vs})=p(\pi \mS|\bar{\vs})$ for any of the $K!$ permutations $\pi \in \mathcal{P}$ of the rows of $\mS$.

\subsection{Permutation Equivariant Flow Matching}
\label{sec:pet_fm}

We want to exploit the permutation invariance of the target distribution, i.e. $p(\mS|\bar{\vs})=p(\pi \mS|\bar{\vs})$.
An equivariant flow matching method relying on optimal transport (OT) has been developed in \cite{klein2024equivariant} to address related problems.
However, it does not apply directly to our setup as the conditioning $\bar{\vs}$ is sample dependent.
Following \cite{chemseddine2024conditional,kerrigan2024dynamic} which extends OT flow matching \cite{tong2024} to the sample dependent conditioning case, we can extend equivariant flow matching as follows. 
\begin{itemize}[leftmargin=*]
\item Sample batches of sources $\vx^{1:B}_1=\mS^{1:B}$ and let $\vc^{1:B}_1=\bar{\vs}^{1:B}$. Then sample $\vx^{1:B }_0$ using \eqref{eq:p(x_0|c)} and let $\vc^{1:B}_0=\vc^{1:B}_1$.
\item Compute $\mC_{i,j}=\min_{\pi} ||\vx^i_0- \pi \vx^j_1||^2+\beta ||\vc^i_0-\vc^j_1||^2$.
\item Solve the discrete OT problem with cost matrix $\mC=(\mC_{i,j})$.
\end{itemize}
We need to pick $\beta \gg 1$ to ensure that we only ``match" samples in the batches that have similar conditioning variables.
Denoting $\pi^{\text{EUC}}_{\vx_0,\vx_1}=\argmin_{\pi} ||\vx_0- \pi \vx_1||$, we minimize the loss
\begin{equation}\label{eq:EquivariantFMloss}
    \mathcal{L}(\theta)=\int_0^1 \mathbb{E}[||v^\theta(t,\vx_t,\bar{\vs})-(\pi^{\text{EUC}}_{\vx_0,\vx_1} \vx_1-\vx_0)||^2]\,\rmd t,
\end{equation}
where $\vx_t=\vx_t(\vx_0, \pi_{(\vx_0,\vx_1)} \vx_1)$ and the expectation is with respect to the joint distribution induced by the OT procedure.
In practice, we found that this procedure, albeit principled, is ineffective in our high-dimensional scenario: the off-diagonal coefficients of $\mC$ are much larger than its diagonal elements for reasonable batch sizes.
So a simple approximation is to use \eqref{eq:EquivariantFMloss} where the expectation is with respect to the joint distribution of $p(\vc)p(\vx_0|\vc)p(\vx_1|\vc)$.
However, this vanilla equivariant flow loss \eqref{eq:EquivariantFMloss} does not work well in practice.
We conjecture that this is partly because this Euclidean distance to a noise signal is not informative for discriminating between sources.

Instead, we leverage permutation invariant training~\cite{kolbaekMultitalkerSpeechSeparation2017} at $t=0$ when the network input contains no information about the preferred permutation of the sources.
First, let the sample-wise loss be
\begin{equation}
   \calL_{\vx_0,\vx_1}(\pi,t) =  \| v^{\theta}(t, \vx_t(\vx_0,\pi\vx_1),\bar{\vs}) - (\pi\vx_1 - \vx_0)\|^2,
    \label{eq:permutedloss}
\end{equation}
and $\pi^{\text{PIT}}_{\vx_0,\vx_1}=\underset{\pi \in \calP}{\arg\min}\ \calL_{\vx_0,\vx_1}(\pi, 0)$.
Then, we propose the following permutation equivariant training (PET) loss,
\begin{align}
    \calL^{\text{PET}}(\theta) = \int_0^1 \lambda_t \E\left[\calL_{\vx_0,\vx_1}(\pi^{\text{PIT}}_{\vx_0,\vx_1}, t)\right]\, \rmd t,
\end{align}
for some positive weight function $\lambda_t$.
Over time, the network will learn to assign particular sources to the permutation leading to the lowest loss.
Then, the particular assignment of the sources is fixed and used to train the loss at a different $t > 0$.
We consider two choices of $\lambda_t$.
The first is $\lambda_t = \frac{1 + \delta_t}{2}$, $\delta_t$ is the delta Dirac function located at $0$.
This trains on $t=0$ and $t\sim\calU[0,1]$ in equal proportions.
The second one is $\lambda_t = \frac{C}{t (1-t)} + \epsilon \delta_t$, with $C,\epsilon>0$ constant, which gives uniform weights in the log signal-to-noise ratio (SNR) domain defined as $r = 20\log_{10}(t/(1-t))$.
This puts more weight on training close to $t=0$ and $1$ where predicting $\vx_1 - \vx_0$ is hardest.
We implement it by choosing $t=(1 + 10^{-r/20})^{-1}$, $r\sim\calU[-80, 100]$ with probability 0.99, or $t=0$ otherwise.
The $\delta_t$ ensures $t=0$ is seen during training.

\textbf{Normalized and Decibel-valued Loss.}
The loss of \eqref{eq:permutedloss} is sensitive to the power of the target $\vx_1 - \vx_0$.
This has the undesirable effect to learn less from smaller amplitude examples.
To remedy this, we propose an inverse SNR loss
\begin{align}
    \calL^{\operatorname{N}}_{\vx_0,\vx_1}(\pi, t) = {\frac{\| v^{\theta}(t, \vx_t(\vx_0,\pi\vx_1 ),\bar{\vs}) - (\pi\vx_1 - \vx_0)\|^2}{\|\vx_1 - \vx_0\|^2}}.
    \label{eq:normalizedpermutedloss}
\end{align}
Furthermore, deviating from the theory of flow matching, we found that training on a decibel valued version of \eqref{eq:normalizedpermutedloss} leads to significant improvement of the separation performance
\begin{align}
    \calL_{\vx_0,\vx_1}^{\operatorname{dB}}(\pi, t) = 10\log_{10} \calL_{\vx_0,\vx_1}^{\operatorname{N}}(\pi, t).
    \label{eq:dbnormalizedpermutedloss}
\end{align}
Such log losses are known to be effective for source separation~\cite{luo_conv-tasnet_2019}.

\begin{figure}
    \centering
    \includegraphics[width=\linewidth]{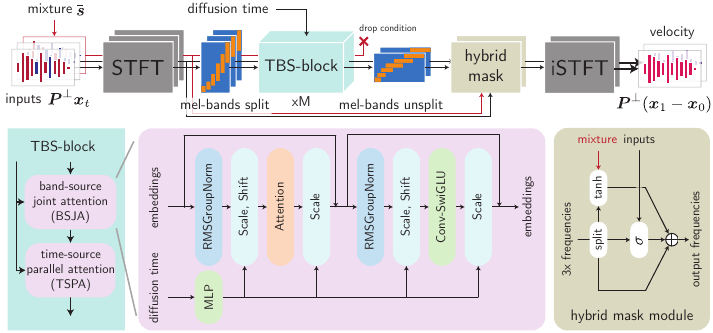}
    \caption{The overall architecture of the network.}
    \label{fig:architecture_overall}
\end{figure}

\subsection{Permutation Equivariant Network}

The final ingredient of equivariant flow matching~\cite{klein2024equivariant} is a permutation equivariant network $v^\theta(t, \vx,\bar{\vs})$, satisfying the property
\begin{align}
   v^\theta(t, \pi \vx,\bar{\vs}) = \pi v^\theta(t, \vx,\bar{\vs}),\quad \forall \pi \in \calP,
\end{align}
i.e., a permutation of the inputs similarly permutes the outputs of the network.
We implement this by restricting equivariance-breaking operations to be applied in parallel to each source.
All interactions between the sources are implemented with a multi-head self-attention (MHSA) module \emph{without} positional encoding.
Fig.~\ref{fig:architecture_overall} shows the general network architecture.
It uses an encoder/decoder architecture with a dual-path transformer pipeline.
The encoder is composed of an STFT, complex magnitude compression, and a learnable Mel-band split layer~\cite{Wang2023-mc}.
The decoder consists of the corresponding inverse operations.
The encoder is followed by global normalization~\cite{luo_conv-tasnet_2019}.
At that stage, the data tensor has four dimensions: time ($T$), band ($B$), source ($S$), and features ($O$).

The main pipeline combines elements of TF-Locoformer~\cite{Saijo2024-rq} (convolutional MLPs) and Diffusion Transformer (DiT)~\cite{Peebles_2023_ICCV} (conditioning on $t$).
The conditioning mixture $\bar{\vs}$ is added to the network as an additional source.
We use a special positional encoding to differentiate it from the sources in the permutation equivariant attention blocks.

\begin{figure}
    \centering
    \includegraphics{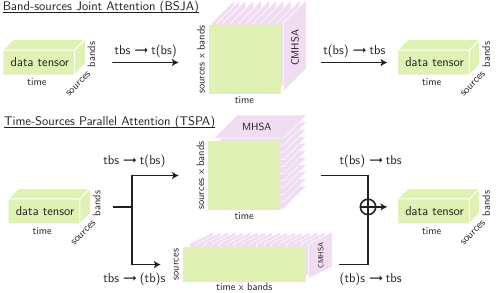}
    \caption{The various attention blocks of the permutation equivariant network.}
    \label{fig:architecture_attention}
\end{figure}

The transformer blocks have a dual path structure where a band-source joint attention (BSJA, Fig.~\ref{fig:architecture_attention} top) and time-source parallel attention (TSPA, Fig.~\ref{fig:architecture_attention} bottom, inspired by a similar architecture in \cite{Dutordoir2022-xe}) are applied alternatively.
Both of these blocks have otherwise a similar structure where the embedding are normalized by RMSGroupNorm~\cite{Saijo2024-rq} and the attention is followed by special MLPs that use convolutional projections along time or bands, respectively, and a gated linear unit non-linearity with a swish gate (Conv-SwishGLU~\cite{Saijo2024-rq}).
Due to the limited interaction between sources, we introduced a convolutional multi-head self-attention (CMHSA).
It has the same structure as MHSA, but the dense projection layers are replaced by convolution operations along the time and band (TSPA only) dimensions.

Because the target can be partially parameterized as a function of the inputs, $\mP^\perp \vx_t$ and $\bar{\vs}$, we introduce a hybrid mask module.
The output of the band unsplit module is projected to 3 times its size and split into mapping, input mask, and mixture mask.
The two masks are used as weights to combine the mapping part with $\mP^\perp\vx_t$ and $\bar{\vs}$.

\subsection{Noise Shaping}
\label{sec:noise_shaping}

The covariance $\mSigma_{\mZ}$ of the noise $\mZ$ in \eqref{eq:p(x_0|c)} is a design parameter. In the simplest case, one can set $\mSigma_{\mZ} = \sigma_0^2 \mI_{KL}$. However, the mixture $\bar{\vs}$ contains prior information about the sources that can be exploited; e.g., at times when the mixture is equal to zero, sources are highly likely to be inactive. We define $\mSigma_{\mZ}$ via a function $\mT\,:\,\R^L\to\R^{L\times L}$ that maps $\bar{\vs}$ to a linear operator, and let $\mZ = \tilde{\mZ} \mT(\bar{\vs})$, for $\tilde{\mZ} \sim \calN(\vzero, \mI_{KL})$.
Since the energy of $\bar{\vs}$ is highly correlated to that of the sources, we can use this to set the power of the added noise.
We explore two choices of $\mT(\bar{\vs})$.
One is $\mT_{\operatorname{pwr}}(\bar{\vs}) = \sigma_{\operatorname{act}}(\bar{\vs}) \mI_L$, the \emph{active} power of the mixture. We first compute the energy envelope of the mixture $\ve(\bar{\vs})$ with a Hamming window filter.
Then, $\sigma_{\operatorname{act}}(\bar{\vs})$ is obtained by averaging the envelope everywhere it is over a minimum energy threshold.
Our second choice, $\mT_{\operatorname{env}}(\bar{\vs}) = \diag(\ve(\bar{\vs})^{\nicefrac{1}{2}} )$ uses the square root of the envelope to scale the signal.
This prevents injecting noise in inactive areas of the signal.

\section{Experiments}

\begin{table}
    \centering
    \caption{Ablation of FLOSS hyperparameters.     
    Loss functions $\calL$, $\calL^{\operatorname{N}}$, and $\calL^{\operatorname{dB}}$ refer to eqs.~\eqref{eq:permutedloss}, \eqref{eq:normalizedpermutedloss}, and \eqref{eq:dbnormalizedpermutedloss}. $\calL^{\operatorname{dB}}_{\text{EUC}}$ uses eq.~\eqref{eq:dbnormalizedpermutedloss} with $\pi^{\text{EUC}}_{\vx_0,\vx_1}$.     
    Weights $\lambda_t$ are described in Section~\ref{sec:pet_fm}.     
    Symbols $\ve(\bar{\vs})$ and $\sigma_{\text{act}}$ are the mixture envelope and active level (Section~\ref{sec:noise_shaping}).     
    The schedule is linear with 25 steps.     
    }
    
    \label{tab:ablation}
    \begin{tabular}{@{}rlccrrrr@{}}
    \toprule
    \# & Loss & $\lambda_t$ & Noise & SI-SDR &   ESTOI &   POLQA &  OVR\\
    \midrule
    1 & $\calL$                     &  $\frac{99+\delta_t}{100}$       & $\ve(\bar{\vs})$ &     9.80 &   0.769 &    2.08 &         2.88 \\
    2 & $\calL^{\operatorname{N}}$  &         &  &    13.26 &   0.824 &    2.82 &         2.92 \\
    3 & $\calL^{\operatorname{dB}}$ &         &  &    \textbf{18.43} &   \textbf{0.905} &    \textbf{3.67} &         2.97 \\
    4 & $\calL^{\operatorname{dB}}_{\text{EUC}}$ &  &  &16.73 &  0.884 &    3.55 &         2.98 \\
    \midrule
    5 & $\calL^{\operatorname{dB}}$ & $1$                             & $\ve(\bar{\vs})$  &    16.42 &   0.881 &    3.53 &         2.98 \\
    6 &  & $\frac{999+\delta_t}{1000}$      &  &    15.98 &   0.878 &    3.50 &         \textbf{2.99} \\
    7 &  & $\frac{1+\delta_t}{2}$          &   &     17.13 &   0.879 &    3.40 &         2.89 \\
    8 &  & $\frac{C}{t (1-t)}$              &  &    17.55 &   0.887 &    3.46 &         2.90 \\
    \midrule
    9 & $\calL^{\operatorname{dB}}$ & $\frac{C}{t (1-t)}$ & $\sigma_{\operatorname{act}}$ &    17.38 &   0.867 &    3.32 &         2.82 \\
    \bottomrule
    \end{tabular}
\end{table}

\subsection{Baseline Methods}

\textbf{Conv-Tasnet}~\cite{luo_conv-tasnet_2019}, a fully convolutional separation network. We use a reimplementation with a time-frequency encoder/decoder rather than the filter bank used in the original paper.
The STFT frame size is 768 with half overlap, the number of channels is 384, and we use 8 blocks for a total of \num{5}M parameters.
\textbf{MB-Locoformer}, a combination of the Mel-band split encoder/decoder~\cite{Wang2023-mc} and the TF-Locoformer~\cite{Saijo2024-rq} transformer architecture. We use 80 Mel-bands, 6 locoformer blocks, and 192 dimensional embeddings resulting in \num{39}M parameters.
Both Conv-TasNet and MB-Locoformer are trained discriminatively by minimizing the negative SI-SDR~\cite{le_roux_sdrhalf-baked_2019} loss.
\textbf{Diffsep}~\cite{scheibler_diffusion-based_2023}, closest to FLOSS, is a diffusion model based on the SDE~\eqref{eq:diffsep_sde}.
We use a DDPM~\cite{ho_denoising_2020} style sampler specialized for Diffsep.
\textbf{EDSep}~\cite{dong_edsep_2025}, a variation of Diffsep with a different parameterization of the predictor and an improved sampler.
Diffsep and EDSep use the improved noise-conditioned score network (NCSN++, \num{47}M parameters)~\cite{song2020score}.

\subsection{Dataset}
We are targeting here speech separation for high quality applications. For this purpose, we use a large dataset of audiobook samples collected from LibriVox~\cite{lookingtolisten}.
The training set is composed of \num{3888399} training and \num{135617} test English language samples.
The samples are restored to \SI{24}{\kilo\hertz} studio quality with the Miipher model~\cite{miipher} .
During training, the samples are randomly cropped to \SI{5}{\second} length.
Their active power is normalized to a random value in \qtyrange{-29}{-19}{\decibel} (relative to 1.0).
Sources are dynamically paired and summed up to create mixtures with an SNR between \qtyrange{-10}{10}{\decibel}.
The evaluation set is created similarly by choosing 2000 samples from the test portion of the dataset and applying the same processing, resulting in 1000 mixtures.

\subsection{Evaluation metrics}

We evaluate the estimated sources in terms of
scale-invariant signal-to-distortion ratio (SI-SDR)~\cite{LeRoux:2018tq}, 
extended short term objective intelligibility (ETOI)~\cite{jensen_algorithm_2016},
wideband perceptual evaluation of speech quality (PESQ)~\cite{rix_perceptual_2001},
perceptual objective listening quality assessment (POLQA)~\cite{Beerends2013-kb},
and the non-intrusive mean opinion score predictor DNSMOS's overall (OVR)~\cite{reddy_dnsmos_2022}.
Signals are resampled to \SI{16}{\kilo\hertz} for PESQ and DNSMOS.
Higher is better for all metrics.

\subsection{Network details, Training, and Inference}
\label{sec:training_inference}
The proposed network uses an STFT encoder/decoder with frame length of \SI{20}{\milli\second}, half overlap, and a normalized Hamming window.
The STFT magnitudes are compressed with an exponent 0.33.
We use an 80 Mel-band split front-end projecting each band to 192 dimensional feature embeddings.
The CMHSA modules use convolution kernel sizes \num{5} (time) and (\num{5}, \num{3}) (time, frequency) for BSJA and TSPA, respectively.
This network has \num{36}M parameters.
All diffusion networks are trained with the AdamW optimizer for \num{250000} steps with an exponential moving average of the weights with coefficient \num{0.999} and weight decay of \num{0.01}.
The learning rate is ramped up linearly from \num{0.0} to \num{1e-4} over \num{25000} steps and decayed with a cosine schedule until the end of training.
The inference for FLOSS is done with an Euler sampler and a linear 25-steps schedule.
We also experiment with single step, and a custom 5-steps schedule with sizes $\{0.95, 4\cdot 10^{-2}, 9\cdot 10^{-3}, 9\cdot 10^{-4}, 1\cdot10^{-4}\}$ that we found to work well.

\begin{table}
    \centering
    \caption{Comparison to baseline methods. Higher is better for all metrics.
    The number of function evaluations (NFE) equals the steps, except for EDSep that uses two per step.
    The dagger ($\dagger$) indicates a custom schedule (see Section~\ref{sec:training_inference}).}
    \label{tab:results}
    \begin{tabular}{@{}lrrrrrr@{}}
    \toprule
    Method & NFE & SI-SDR &   ESTOI &   PESQ &   POLQA &   OVR \\ %
    \midrule
    Conv-TasNet & 1 & 12.50 &  0.819 &  2.57 &   2.46 &   2.78 \\
    MB-Locoformer    &  1 & 17.97 &  0.903 &  3.25 &  3.79 &  2.23 \\
    \midrule
    Diffsep    & 25 &  7.49 &  0.760 &   2.23 &  2.32 &  2.87 \\
    EDSep      & 50 & 11.45 &  0.792 &   2.44 &  2.62 &  2.77 \\
    \midrule
    FLOSS (\#3)& 25 & 18.43 &   \textbf{0.905} &  3.36 &  3.67 & 2.97 \\
               & 5$^\dagger$  & \textbf{19.13} &   0.903 &  \textbf{3.45} &  \textbf{3.86} & \textbf{2.99} \\
               & 1  & 19.12 &   0.900 &  3.36 &  3.80 & 2.94 \\
    \bottomrule
    \end{tabular}
\end{table}

\subsection{Results}

Table~\ref{tab:ablation} shows an ablation of various parts of the proposed algorithms.
Rows \#1--3 show that the normalized (\#2) and log domain (\#3) loss functions bring large improvements to all metrics.
In rows \#3--4, we observe that PIT outperforms Euclidean distance to decide the source permutation.
Rows \#3 and \#5--8 compare the weight functions.
The weight of $\delta_t$ is important and $0.01$ (\#1) performs best.
Row \#8 shows the SNR uniform schedule that leads to competitve intrusive metrics, but hurts the DNSMOS (OVR) somewhat.
Finally, mixture shaped (\#8, $\ve(\bar{\vs})$) does better than constant (\#9, $\sigma_{\text{act}}$) noise on all metrics.

Table~\ref{tab:results} compares FLOSS to baseline predictive and generative separation methods.
For this, we use the best model of Table~\ref{tab:ablation} (\#3), which trains 1\% of examples at $t=0$.
FLOSS performs best among all methods.
Not only does it score well on DNSMOS metrics, indicating naturalness, it also tops the intelligibility metrics, and even the sample level metrics such as SI-SDR, with strong one-step performance.     
The predictively trained MB-Locoformer scores well on intrusive metrics, but markedly lower on DNSMOS, which is confirmed by informal listening revealing some regression to the mean artifacts.
While the performance of Diffsep compared to Conv-TasNet is in line with the results reported in~\cite{scheibler_diffusion-based_2023}, it does worse amongst all methods despite using a similar training loss and the same network.
We can attribute this to the fact that its design does not preserve mixture consistency.
Speech samples are available as supplementary material.

\section{Conclusion}
We have proposed FLOSS, a flow matching-based method for audio source separation. It incorporates mixture consistency by design.
Furthermore, it handles the permutation ambiguity of sources in a principled way.
We design a specialized source permutation equivariant network architecture that combines building blocks of state-of-the-art audio separation networks.
Its band-split architecture makes it suitable for handling large sampling rates for high-quality audio processing.
The experiments demonstrate a large improvement in performance compared to Diffsep, the most similar method for generative audio source separation.
In the future, we would like to extend the method to tackle noisy and distorted mixtures, broadening its applicability.
\newpage
\bibliographystyle{IEEEtran}
\bibliography{refs}

\end{document}